\documentclass[preprintnumbers,aps,pre,twocolumn,superscriptaddress,showpacs,amsmath,amssymb]{revtex4-1}

\usepackage{color}
\usepackage{graphicx}
\usepackage{dcolumn}
\usepackage{bm}

\newcommand{\nvec}{\mathbf{n}}
\newcommand{\lvec}{\mathbf{l}}
\newcommand{\mvec}{\mathbf{m}}

\newcommand{\Qvec}{\mathbf{Q}}
\newcommand{\Qij}{Q_{ij}}
\newcommand{\tildeQij}{\tilde Q_{ij}}
\newcommand{\Acal}{\mathcal{A}}
\newcommand{\Fcal}{\mathcal{F}}
\newcommand{\Scal}{\mathcal{S}}

\newcommand{\dd}{\mathrm{d}}

\DeclareMathOperator{\Tr}{Tr}

\begin{document}

\title{Filling and wetting transitions of nematic liquid crystals on
  sinusoidal substrates}

\author{P. Patr\'icio}
\email[]{patricio@cii.fc.ul.pt}
\affiliation{Instituto Superior de Engenharia de Lisboa,
Rua Conselheiro Em\'idio Navarro 1, P-1959-007 Lisboa, Portugal}
\affiliation{Centro de F{\'\i}sica Te\'orica e Computacional,
Universidade de Lisboa,
Avenida Professor Gama Pinto 2, P-1649-003 Lisboa Codex, Portugal}

\author{N. M. Silvestre}
\affiliation{Centro de F{\'\i}sica Te\'orica e Computacional,
Universidade de Lisboa,
Avenida Professor Gama Pinto 2, P-1649-003 Lisboa Codex, Portugal}

\author{C.-T. Pham} \affiliation{Laboratoire d'Informatique pour la
  M\'ecanique et les Sciences de l'Ing\'enieur, CNRS-UPR 3251,
  Universit\'e Paris-Sud 11, BP 133, F-91403 Orsay Cedex, France }

\author{J. M. Romero-Enrique} \affiliation{Departamento de F\'\i sica
  At\'omica, Molecular y Nuclear, Area de F\'\i sica Te\'orica
  Universidad de Sevilla, Apartado de Correos 1065, 41080 Sevilla,
  Spain }%

\date{May 2011}

\begin{abstract}

  Close to sinusoidal substrates, simple fluids may undergo a filling
  transition, in which the fluid passes from a dry to a filled state,
  where the interface remains unbent but bound to the
  substrate. Increasing the surface field, the interface unbinds and a
  wetting transition occurs.  We show that this double-transition
  sequence may be strongly modified in the case of ordered fluids, such
  as nematic liquid crystals.  Depending on the preferred orientation
  of the nematic molecules at the structured substrate and at the
  isotropic-nematic interface, the filling transition may not exist,
  and the fluid passes directly from a dry to a complete-wet state,
  with the interface far from the substrate.  More interestingly, in
  other situations, the complete wetting transition may be prevented,
  and the fluid passes from a dry to a filled state, and remains in this
  configuration, with the interface always attached to the substrate,
  even for very large surface fields. Both transitions are only
  observed for a same substrate in a narrow range of amplitudes.

\end{abstract}

\maketitle

\section{Introduction}

Wetting on smoothly structured substrates show a rich phenomenology.
For simple fluids,
the wetting behavior on structured substrates
presents a variety of phenomena
\cite{Rascon_2000_2,Callies_Quere_2005,Bonn_etal_2009}, for which
several physical laws were theoretically proposed, as the
\textit{Wenzel} \cite{Wenzel_1936} or the \textit{Cassie-Baxter laws}
\cite{Cassie_Baxter_1944}. These laws extend \textit{Young's law}
\cite{deGennes_1985} for planar substrates, including new substrate
geometrical parameters.  If these laws are justified in some
situations \cite{Patricio_Pham_RomeroEnrique_2008}, they may also fail
in general theoretical or experimental conditions \cite{Swain_1998}.
In addition to the usual wetting transition, simple fluids at bulk coexistence
may present other surface transitions as the filling transition
\cite{Rascon_1999,Rejmer_2000}.
From a thermodynamical point of view, the filling transition in simple fluids
always precedes a complete wetting transition. However, the filling
transition may not exist for shallow substrates \cite{Rascon_1999},
but if it does exist, it always precedes the wetting transition.

The existence of long-range order in complex fluids may alter the
scenario depicted above \cite{Patricio_Romero_etal_2011}.
In a nematic liquid crystal, there is an orientational (continuous)
long-range order. However, the presence of substrates which favor
specific orientations frustrate the liquid crystal tendency to align
along a given direction. As a consequence, elastic distortions emerge,
altering the subtle balance between the different free-energy
contributions which lead to the surface transitions on structured substrates.

In this paper, we show that the filling-wetting sequence
which is observed in simple fluids on sinusoidal substrates
may be deeply modified
when we have a nematic liquid crystal instead of a simple fluid.
Depending on the physical elastic parameters of the nematic, several
scenarios may occur: we may only have a wetting transition; or a
filling transition only, preventing the subsequent complete wetting
transition; or the double sequence as in the simple fluid case. Both
filling and wetting transitions are only observed for the same
geometry in a narrow range of amplitudes.

This article is organized as follows: in Section II.A. we briefly
review the main simple analytical results about filling and wetting
transitions of simple fluids.  In Section II.B., we extend these
results to nematic liquid crystal systems.  In Section III, we use the
Landau-de Gennes model of nematics to study their wetting behavior on
a sinusoidal substrate. At the end, we draw some conclusions.

\section{Macroscopic analysis}

\subsection{Thermodynamic description of filling and wetting
  transitions for simple fluids}

\begin{figure}[t]
\centerline{\includegraphics[width=.95\columnwidth]{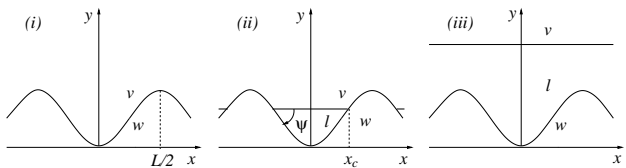}}
\caption {Schematic representation of (i) dry, (ii) filled and (iii)
  wet states of a simple fluid (l) on a sinusoidal substrate (w) of
  wave length $L$ with vapor (v) above.}
\label{fig1}
\end{figure}

In this section, we review the simple description of filling and
wetting transitions for simple fluids, following the ideas presented
in \cite{Rejmer_2000}.  To that end, let us consider a structured
substrate, like the one presented in Fig. \ref{fig1}.  The substrate
is translationally invariant along $z$, with $L_z$ being the total
length along this direction, and periodic along the $x$-axis with a
period $L$. The substrate profile can be described by a simple even
function $g(x)$.  To simplify the analysis, we consider a
monotonically increasing function for $0<x<L/2$.

We suppose the substrate favors the liquid phase and consider three
possible interfacial states: i) dry, ii) filled and iii) wet (see
Fig. \ref{fig1}).  If the bulk free energies of the vapor and liquid
phases are the same (so they can be set equal to zero), and if we
neglect interactions between the interfaces, then the energies
associated with each state will be simply the sum of surface energy
terms. Let us denote $\Scal$ as the whole substrate area, $\Acal$ its
projection on the $x-z$ plane and $\Fcal$ the excess interfacial free
energy.  Then, the free energy of the dry state (see
Fig. \ref{fig1}(i)) is
\begin{equation}
\Fcal^{(i)}=\Scal\sigma_{vw}
\end{equation}
where $\sigma_{vw}$ is the vapor-wall (substrate) surface tension.

The free energy corresponding to the filled interfacial state (see
Fig. \ref{fig1}(ii)) is
\begin{equation}
  \Fcal^{(ii)}=S(x_c)\sigma_{lw}+(\Scal-S(x_c))\sigma_{vw}+\left(
    \frac{2x_c}{L}\right)\Acal\, \sigma_{lv}
\label{free-energy-filled}
\end{equation}
where $\sigma_{lw}$ is the liquid-wall (substrate) surface tension,
$\sigma_{lv}$ is the liquid-vapor surface tension, and $S(x_c)$ is the
substrate area in contact with the liquid phase, which can be obtained
in terms of the abscissa $x_c$ of the contact line of the filled
region with the substrate as:
\begin{equation}
  S(x_c)=\Scal\frac{\int^{x_c}_{-{x_c}}\sqrt{1+g_x^2}dx}{\int^{L/2}_{-L/2}\sqrt{1+g_x^2}dx}
\end{equation}
where $g_x$ is the derivative of $g$ with respect to $x$. The value of
$x_c$ can be obtained by minimisation of the free energy
Eq. (\ref{free-energy-filled}) with respect to it. Making use of
Young's law, which relates the surface tensions mentioned above with
the contact angle of a sessile drop of the liquid in a planar
substrate, $\theta_\pi$,
\begin{equation}
\sigma_{vw}-\sigma_{lw}=\sigma_{lv}\cos\theta_\pi
\label{Young}
\end{equation}
it can be shown that the minimum energy condition is
fulfilled when $x_c$ satisfies \cite{Rejmer_2000}:
\begin{eqnarray}
0&=&2\sigma_{lv}(1-\sqrt{1+g_x^ 2(x_c)}\cos\theta_\pi)\nonumber\\
&=&2\sigma_{lv}\left(1-\frac{\cos\theta_\pi}{\cos\psi}\right)
\end{eqnarray}
showing that the filling region must make contact with the substrate
with an angle $\psi=\theta_\pi$. For smooth substrates, this solution
is a local minimum only if $g_{xx}(x_c)=d^2 g/dx^2(x_c)<0$.

Finally, the free energy for the wet state (see Fig. \ref{fig1}(iii))
is given by:
\begin{equation}
\Fcal^{(iii)}=\Scal\sigma_{lw}+\Acal\sigma_{lv}
\end{equation}

Different transitions between the different interfacial states can be observed.
The transition from the dry state directly to the wet state
is possible when the energies of these configurations are the same, that is,
\begin{equation}
\Fcal^{(iii)}-\Fcal^{(i)}=0=\Scal(\sigma_{lw}-\sigma_{vw})+\Acal\sigma_{lv}
\label{df31}
\end{equation}
Using Young's law Eq. (\ref{Young}), it is possible to deduce from
Eq. (\ref{df31}) Wenzel's law for the dry-to-wet transition:
\begin{equation}
\frac{\Scal}{\Acal}\cos\theta_\pi=r\cos\theta_\pi=\cos\theta_r=1
\label{wenzel}
\end{equation}
where $r=\Scal/\Acal>1$ is the roughness of the substrate.  As
$\theta_r\to 0$, a wetting transition may occur. However, this
transition may be preempted by the presence of filled states.  The
filling or unbending transition between a dry and a filled state may
occur if the free energies of these interfacial states are the same:
\begin{equation}
  \Fcal^{(ii)}-\Fcal^{(i)}=0=S(x_c)(\sigma_{lw}-\sigma_{vw})+\Acal \left(\frac{2x_c}{L}\right)\sigma_{lv}
  \label{df21}
\end{equation}
which leads to the equation
\begin{equation}
  \frac{\int^{x_c}_{-{x_c}}\sqrt{1+g_x^2}dx}{2x_c}\cos\theta_\pi=r_c\cos\theta_\pi=1
\end{equation}
where $r_c>1$. If this transition occurs at a lower temperature than
the wetting transition predicted by the Wenzel law (i.e. for larger
values of $\theta_\pi$, as $\theta_\pi$ usually decreases with
temperature), then a filling transition occurs and Wenzel's law does
not hold.  Under these conditions, macroscopics dictates that the
wetting transition between a filled and a wet interfacial state must
occur under the same conditions as in the planar substrate
(i.e. $\theta_\pi=0$). However, this prediction can be changed as
interactions between the substrate and the liquid-vapor interface are
taken into account. So, if there is a first-order wetting transition
for the planar substrate, the wetting transition is still first-order
but shifted to larger values of $\theta_\pi$ (smaller
temperatures)\cite{Rejmer_2000}.  On the other hand, for planar
continuous wetting its location and nature is unchanged
\cite{Rascon_1999}.

\subsection{Thermodynamic description of filling and wetting
  transitions for nematic fluids}

\begin{figure}[t]
\centerline{\includegraphics[width=.7\columnwidth]{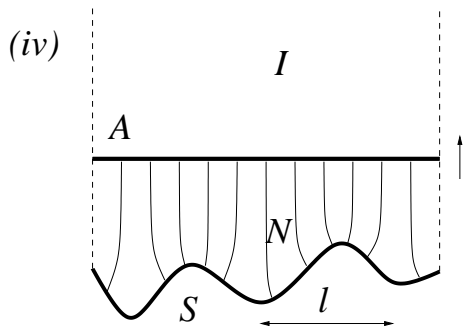}}
\caption
{Schematic representation of the nematic wet state on a rough substrate
of typical $l$ at isotropic-nematic phase coexistence.}
\label{fig2}
\end{figure}
If instead of a simple fluid under saturation conditions, we have a
liquid crystal at the nematic-isotropic phase coexistence, some new
phenomena will inevitably arise.  Let us consider that the substrate favors
the nematic phase at a particular anchoring condition.  From a
thermodynamical point of view, if the bulk free energies of the phases
are the same, the total energy of the nematic interfacial state has a
new positive extra term $\Fcal_d>0$, corresponding to the Frank
elastic energy of the nematic distortions imposed by the anchoring at
the substrate.  Fig.\ref{fig2} shows schematically an example of a
nematic phase in contact with a substrate with homeotropic anchoring
conditions (however, the theoretical arguments presented here are very
general, and independent on the particular anchoring conditions
involved). The total free energy of this interfacial state can be
written as
\begin{equation}
\Fcal^{(iv)}=\Scal\sigma_{nw}+\Acal\sigma_{in}+\Fcal_d
\end{equation}
The elastic deformation creates an effective long-range repulsion
between the substrate and the interface: the closer the interface is
to the substrate, the more constrained the order will be, leading to
higher energies.  As the interface goes to infinity, the system will
only have a characteristic length scale, that is imposed by the
substrate (see Fig. \ref{fig2}).  If no defect arises, the elastic
contribution may be simply estimated as $\Fcal_d=(K\Scal/l)\tilde
\Fcal_d$, where $K$ is the Frank elastic constant, and $\tilde
\Fcal_d$ is a reduced elastic contribution to the interfacial free
energy, only dependent on the substrate geometry, but not on its scale
$l$ \cite{Berreman_1972}.  In some cases, topological defects nucleate
close to or on the substrate.  The presence of defects imposed by the
substrate introduce a $\ln(l/\xi)$ dependence on the reduced elastic
contribution $\tilde \Fcal_d$
\cite{RomeroEnrique_Pham_Patricio_2010}.  This dependence is easily
justified. If the defects are induced by the substrate, they will
place themselves at a distance of order of $l$ (the characteristic
length of the substrate), usually close to its crests and/or
troughs. On the other hand, the cut-off length is given by the
size of the defects, which scales with the correlation length $\xi$.

The dry-to-nematic wet transition may be derived from the generalized
Wenzel law \cite{Patricio_Pham_RomeroEnrique_2008}:
\begin{equation*}
\Fcal^{(iv)}-\Fcal^{(i)}=0=
  \Scal(\sigma_{nw}-\sigma_{iw})+\Acal\sigma_{in}+\Scal\frac{K}{l}\tilde \Fcal_d 
\end{equation*}
that can be rewritten as
\begin{equation}
r\cos\theta_\pi=1+\frac{rK\tilde \Fcal_d}{l\sigma_{in}}.
\label{generalized_wenzel_law}
\end{equation}
For large enough substrate lengths, $l\gg K/\sigma_{in}$, the effects
of elastic terms are unimportant, and Wenzel's law is recovered.
Usually, $K$ and $\sigma_{in}$ are not independent, and their ratio
scales with the correlation length $\xi$.  In fact, using Landau-de
Gennes model for nematic liquid crystals (see next section), it is
possible to show that $K/\sigma_{in}\sim 10 \xi$ (at coexistence
temperature).

For the simple fluid case ($\cos\theta_\pi=1/r<1$), the dry-to-wet
transition for the rough substrate occurs always before the wetting
transition for the planar substrate, as the surface field or the
temperature is increased.  In contrast, for the nematic case, the
wetting transition for a rough substrate may occur either before (if
the roughness is more important) or after the wetting transition for
the planar substrate (if the elastic deformations are more
important).  These deviations with respect to Wenzel's law are enhanced
if defects are present, as the elastic term of the generalized
Wenzel's law (Eq. (\ref{generalized_wenzel_law})) will decay slower
with increasing $l$.  The sawtoothed substrate exemplifies this effect
\cite{RomeroEnrique_Pham_Patricio_2010}.

We now turn to the case of nematic filled states.  Once again, it is
useful to first consider configurations without defects, which do not
appear if the substrate is shallow enough. Given the fact that the nematic-isotropic
interface favors a particular nematic anchoring \cite{deGennes_1970},
if the interface is near the substrate, it will induce strong elastic
deformations, because the nematic director is constrained to follow
the anchoring conditions on both the interface and the substrate. Thus the
elastic energy of the deformations may be large enough to prevent any
filling transition, even if it would normally
occur in the case of a simple fluid.  It should also be pointed out that the interfacial free
energies of the filled and wet states are normally very close to each
other \cite{Rejmer_2000}, and a very tiny perturbation can exchange
their relative stability.

If the substrate imposes defects on the nematic matrix, a different
picture emerges.  The case of a sinusoidal grating is studied in
detail in the next section.  If the substrate is rough enough, and if
the nematic-isotropic interface favors for example homogeneous planar
anchoring, then the filled state (with the interface placed in
between the crests of the sinusoidal substrate), may always have a
smaller energy than the wet state for in the latter case, a periodic array
of defects is necessarily created at the top of each crest.


\section{The sinusoidal substrate. Numerical results}

In this section, in order to substantiate our arguments, let us
consider a nematic liquid crystal in contact with a sinusoidal
grating, characterized by a wavevector $q=2\pi/L$, and an amplitude
$A$ (see also Fig. \ref{fig1}):
\begin{equation}
  g(x)=A(1-\cos qx)
\end{equation}
At the substrate, the nematic molecules preferentially align
homeotropically, i.e.  perpendicularly to the substrate. The system is
translationally invariant along the out-of-plane axis $z$ and periodic
along the $x$ axis.  Finally we impose that, far from the substrate,
we have an isotropic ordering.

\subsection{The Landau-de Gennes model}

In the Landau-de Gennes (LdG) model, both isotropic and nematic phases can
be locally represented by a traceless, symmetric order-parameter tensor with
components:
\begin{equation}
\Qij = \frac 3 2 S[n_in_j - \frac 1 3 \delta_{ij}] + \frac 1 2 B
[l_il_j - m_i m_j],
\label{tensor}
\end{equation}
where $n_i$ are the Cartesian components of the director field
$\nvec$, $S$ is the nematic order parameter which measures the
orientational ordering along the nematic director, and $B$ is the
biaxiality parameter, which measures the ordering of the molecules on
the orientations perpendicular to $\nvec$, characterized by the
eigenvectors $\lvec$ and $\mvec$. In our model, we will only consider
in-plane de- formations, although out-of-plane or twist deformations
may also be important (a twist instability may occur for particular
choices of parameters \cite{Patricio_Telo_Dietrich_2002}).  In this
case, $\nvec = (\cos\theta,\sin\theta,0)$, and the tensor order
parameter has only three independent components (namely $Q_{11}$,
$Q_{22}$, and $Q_{12}$).

The LdG free energy may be written as
\begin{equation}
  \mathcal{\Fcal}_{\mathrm{LdG}} = \int_{\mathcal{V}} (\phi_{\mathrm{bulk}} +
  \phi_{\mathrm{el}})\,\dd V + \int_{\mathcal{S}} \phi_{\mathrm{surf}}\,\dd s
  \label{free_energy}
\end{equation}
where $\phi_{\mathrm{bulk}}$ is the bulk free energy density,
$\phi_{\mathrm{el}}$ is the elastic free energy density, and
$\phi_{\mathrm{surf}}$ is the surface free energy, defined
as~\cite{deGennes_Prost_1995}:
\begin{align}
  & \phi_{\mathrm{bulk}} = a \Tr \Qvec^2 - b \Tr \Qvec^3 + c [\Tr
  \Qvec^2]^2&&\\ & \phi_{\mathrm{el}} =
\frac{L_1}{2}\partial_k \Qij\partial_k \Qij + \frac{L_2}{2} \partial_j \Qij
\partial_k Q_{ik} &&\\ & \phi_{\mathrm{surf}} = - \frac 2 3 w \Tr
[\Qvec\cdot\Qvec_{\mathrm{surf}}]
\end{align}
where $a$ depends linearly on the temperature, $b$ and $c$ are
positive constants, and $L_1$ and $L_2$ are positive parameters
related to the elastic constants. If we rescale all the variables as
follows \cite{Andrienko_Tasinkevytch_Patricio_etal_2004}: $\tilde
\Qvec=6c\Qvec/b$, the positions $\tilde{\mathbf{r}}={\mathbf{r}}/\xi$,
where the correlation length $\xi$ is defined as $\xi^2=8 c (3 L_1+
2L_2)/b^2$, and $\mathcal{\tilde \Fcal}_{\mathrm{LdG}} = 24^2 c^3
\mathcal{\Fcal}_{\mathrm{LdG}}/\xi^3 b^4$, we get $\mathcal{\tilde
  \Fcal}_{\mathrm{LdG}} = \int_{\mathcal{\tilde V}} (\tilde
\phi_{\mathrm{bulk}} + \tilde \phi_{\mathrm{el}})\,\dd \tilde V +
\int_{\mathcal{\tilde S}} \tilde \phi_{\mathrm{surf}}\,\dd \tilde s$,
with rescaled free energy densities:
\begin{align}
  & \tilde \phi_{\mathrm{bulk}} = \frac 2 3 \tau \Tr \tilde \Qvec^2 -
  \frac 8 3 \Tr \tilde \Qvec^3 + \frac 4 9 [\Tr \tilde \Qvec^2]^2&&\\
  & \tilde \phi_{\mathrm{el}} = \frac {1}{3+2\kappa}[\tilde \partial_k
  \tildeQij\tilde \partial_k \tildeQij + \kappa \tilde \partial_j
  \tildeQij \tilde \partial_k \tilde Q_{ik}] &&\\
  & \tilde \phi_{\mathrm{surf}} = - \frac 2 3 \tilde w \Tr [\tilde
  \Qvec\cdot\tilde \Qvec_{\mathrm{surf}}]
\end{align}
Here $\tau=24ac/b^2$ is a dimensionless temperature, $\kappa=L_2/L_1$
is an elastic dimensionless parameter (for stability reasons, the
elastic parameter is restricted to the values, $\kappa>-3/2$) and
$\tilde w=16wc/b^2 \xi $ is the dimensionless anchoring strength.
Hereafter, we will consider these rescaled expressions, so we will
drop the tilde notation.

In our model, we will place ourselves at coexistence temperature,
$\tau=1$, for which the bulk free-energy density has two minima,
corresponding to $\phi_{\mathrm{bulk}}=0$ for rescaled scalar order
parameters $S_\mathrm{I}=0$ (isotropic phase) and $S_\mathrm{N}=1$
(nematic phase).  It is important to note that the order parameter $S$
in the coexisting nematic phase is rescaled, so its value in real
units is $b/6c$, which must be smaller than 1 (typically $\approx
0.4$).  If the elastic parameter $\kappa$ is positive (negative), the
nematic prefers to align parallel (perpendicular) to a possible
nematic-isotropic interface. Finally, $\Qvec_{\mathrm{surf}}$ defines
the favored tensor at the substrate.  We favor a homeotropic alignment
of the nematic by setting
$\Qvec_{\mathrm{surf}}=(3\boldsymbol{\nu}\otimes
\boldsymbol{\nu}-1)/2$, with $\boldsymbol{\nu}$ the normal vector to
the substrate, establishing a direct connection to previous papers
\cite{Sheng_1976,Sheng_1982,Braun_1996}.

\subsection{Numerical procedure}

\begin{figure}[t]
\centerline{\includegraphics[width=.95\columnwidth]{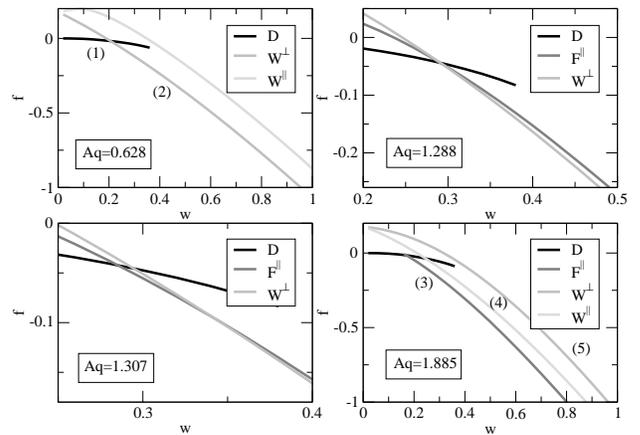}}
\caption
{Plot of the free energy densities (per projected unit area) of different branches of (meta)stable states, as a function of the anchoring strength $w$, for a sinusoidal substrate with $L=10\xi$, $\kappa=2$
and $Aq=0.628$, $1.288$, $1.307$ and $1.885$. Numbers correspond to the typical configurations shown in Fig.\ref{fig4}. See text for explanation.
}
\label{fig3}
\end{figure}
For every set of model parameters, we numerically minimized the
Landau-de Gennes free energy, using a conjugate-gradient method. The
numerical discretization of the continuum problem was performed with a
finite-element method combined with adaptive meshing in order to
resolve the different length scales
\cite{Patricio_Tasinkevych_Telo_2002}.  Due to the translational
symmetry along the $z$ axis and the periodicity on the $x$ axis, we
restricted the minimisation to a section of the system perpendicular
to the $z$ axis and with a width along the $x$ axis equal to the
period $L$ of the substrate. We used periodic boundary conditions on
the lateral sides, and the fluid on the substrate, which corresponds
to the bottom boundary of our unit cell, has not an imposed ordering
or anchoring, although a nematic phase with homeotropic orientation is
energetically favourable. Finally, we impose different fixed boundary
conditions at the cell top (which we place at a large height $H$) in
order to explore the different interfacial states we may observe. The
value of $H$ was varied until we found convergence in the free
energy. In order to find the dry ($D$) or the filling ($F$)
interfacial states, we fix a bulk isotropic phase on top of the cell.
Under these conditions, different states may be obtained when we vary
the initial condition (for example, by considering initial conditions
with different heights for the nematic filling region). The complete
wet ($W$) interfacial states are obtained by imposing a fixed nematic
phase at the cell top with a homogeneous nematic director along either
the $x$ axis ($W^\parallel$) or the $y$ axis ($W^\perp$) (see Fig. \ref{fig4} for illustrations). The free
energy for the wet states is calculated by adding to the numerically
obtained free energy the contribution of an isotropic-nematic
interface parallel to the $x-z$ plane, $L\sigma_{in}$, calculated also
numerically (and in some cases analytically) and corresponding to the
most favoured anchoring conditions (homeotropic to the interface for
$\kappa<0$, planar for $\kappa>0$). In this way we consider the
presence of a nematic-isotropic interface infinitely away from the
substrate in the wet state, allowing us to neglect the elastic
deformations which may exist between the top cell and the
nematic-isotropic interface. The true equilibrium state will be the
state which gives the least free energy at the same thermodynamic
conditions.

\subsection{Numerical results}

\subsubsection{Case of $\kappa >0$}

\begin{figure}[t]
\centerline{\includegraphics[width=.95\columnwidth]{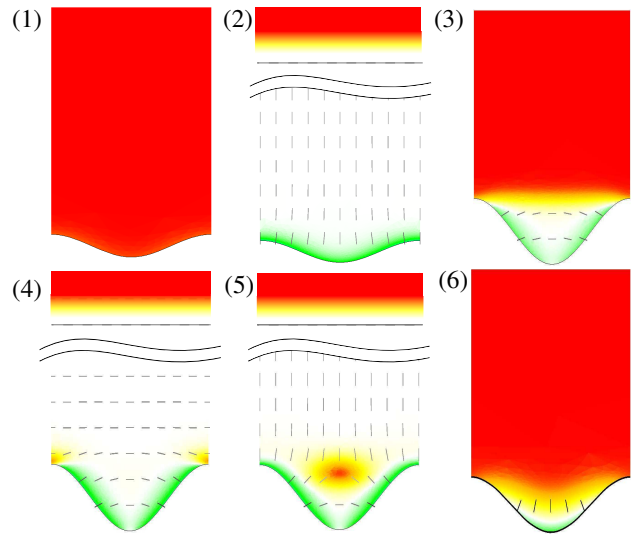}}
\caption {Typical configurations for a sinusoidal substrate with
  $L=10\xi$: from left to right and top to bottom, (1) a $D$ state
  ($\kappa=2$, $Aq=0.628$ and $w=0.2$), (2) a $W^\perp$ state for
  shallow substrates ($\kappa=2$, $Aq=0.628$ and $w=1.0$), (3) a
  $F^\parallel$ state ($\kappa=2$, $Aq=2.199$ and $w=0.4$), (4) a
  $W^\parallel$ state for rough substrates ($\kappa=2$, $Aq=2.199$ and
  $w=1.0$), (5) a $W^\perp$ for rough substrates ($\kappa=2$,
  $Aq=2.199$ and $w=1.0$), and (6) a $F^\perp$ state ($\kappa=-1/2$,
  $Aq=1.257$ and $w=0.4$). Numbers correspond to those shown in
  Fig. \ref{fig3}. (Color online)}
\label{fig4}
\end{figure}

As an example, in Fig. \ref{fig3} we plot the free energy per
projected unit area $f=\Fcal/\Acal$ corresponding to the most relevant
interfacial states for $L=10\xi$, $\kappa=2$ (favoring a parallel
anchoring at the nematic-isotropic interface) and different values of
$Aq=0.628$, $1.288$, $1.307$ and $1.885$. For the shallowest case
$A=\xi$ we plot the branches corresponding to $D$ states (which is the
most stable for small $w$), $W^\perp$ states and $W^\parallel$. No $F$
states are observed.  As it is expected, the $W^\parallel$ branch has
always larger free energy than the others, since for shallow
substrates the elastic deformations are stronger for this state than
for the $W^\perp$ state.  On the other hand, the $D$ and $W^\parallel$
branches intersect at a first-order wetting transition for
$w_W=0.2056$, so $W^\perp$ branch is the most stable for $w>w_W$
(typical configurations for both $D$ and $W^\perp$ states are shown at
Fig.\ref{fig4}-(1) and \ref{fig4}-(2)).

For rougher substrates, the scenario is completely different (see
Fig. \ref{fig3}). For $Aq=1.885$ four different (locally) stable
branches are observed: $D$ states (again the most stable for small
$w$), $W^\perp$ states, $W^\parallel$ states and a filled state where
the nematic is planar at the nematic-isotropic
interface. This new state will be denoted by $F^\parallel$ (a typical
configuration is shown in Fig. \ref{fig4}-(3)). For small $w$, the
nematic-isotropic interface is almost parallel to the $x-z$ plane. As
$w$ increases, the position of the nematic-isotropic interface for the
$F^\parallel$ state is lifted and curved slightly, and it eventually
pins at the crests of the substrate.  Regarding the complete wet
states, the free energy of the $W^\perp$ branch is always larger than
those of the $W^\parallel$ states (unlike for shallow wedges).  Typical
configurations are shown in Fig. \ref{fig4}-(4) ($W^\parallel$) and
\ref{fig4}-(5) ($W^\perp$).  Now roughness induces larger deformations
in $W^\perp$, leading to a higher free energy.  In both cases
topological defects may nucleate either close to the substrate troughs
($W^\perp$ states) or the crests ($W^\parallel$ states). We can see
that the $F^\parallel$ branch has always a smaller energy than the
wetting states, so there is a first-order filling transition between a
$D$ and $F^\parallel$ state at $w_F=0.1693$.  However, we do not
observe a wetting transition, even for large $w$. This latter feature
can be rationalised by the fact that the $F^\parallel$ state does not
present topological defects as does the $W^\parallel$ state.

The crossover from the scenarios described above occurs at
intermediate values of $Aq$. Above some value of $Aq$ the branch of
$F^\parallel$ emerges as a metastable branch. At $Aq\approx 1.288$ a
triple point occurs, since the $D$, $F^\parallel$ and $W^\perp$
branches intersect at the same value of $w=0.291$ (see
Fig. \ref{fig3}).  Above this triple point, the $F^\parallel$ states
branch crosses both the $D$ and $W^\perp$ branches, so filling and
wetting transitions are observed for the same geometry. However, the
value of $w$ for the wetting transition has a steep increase with
$Aq$, so the wetting transition is no longer observable for $Aq$
larger than $1.33$.

\begin{figure}[t]
  \centerline{\includegraphics[width=.95\columnwidth]{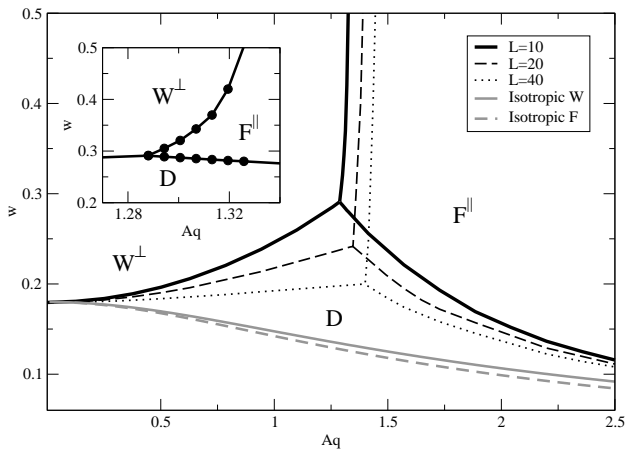}}
  \caption {Global phase diagram for nematic adsorption on sinusoidal
    substrates with $\kappa=2$. Black lines correspond the phase boundaries
    between the $D$, $W^\perp$ and $F^\parallel$ states for $L=10\xi$
    (solid lines), $L=20\xi$ (dashed lines) and $L=40\xi$ (dotted
    lines). Gray lines correspond to the transition from dry to filled state (dashed line) and from filled to wet state (solid line) in the case of isotropic fluid. Inset: zoom of the phase boundaries around the triple
    point for $L=10\xi$.}
\label{fig5}
\end{figure}

The global adsorption phase diagram for $\kappa=2$ is shown in
Fig. \ref{fig5}. The domains for the different possible stable phases
($D$, $W^\perp$ or $F^\parallel$) are divided by first-order
transition lines which meet at a triple point. Note that the
$F^\parallel$-$W^\perp$ transition line is almost vertical, so the
range in which both filling and wetting transitions are observed is
quite narrow.  As the substrate wavelength $L$ increases, the
importance of the elastic term of the nematic free energy diminishes,
and the $D$-$W^\perp$ and $D$-$F^\parallel$ transition lines shift
towards lower values of $w$, approaching the values obtained for simple
fluids within the macroscopic theory outlined in Section II.  However, due to the presence of elastic distortions,
larger deviations than for simple fluids can  be observed in all the cases,

For the
$D$-$W^\perp$ wetting transition we can estimate
this deviation through the modified Wenzel law prediction,
Eq. (\ref{generalized_wenzel_law}). The different terms in this expression can
either be evaluated exactly or estimated numerically. Firstly, the nematic
order parameter for the LdG model close to a planar substrate in the presence
of a bulk nematic or isotropic phase can
be solved analytically \cite{RomeroEnrique_Pham_Patricio_2010}. The exact
solutions of the corresponding Euler-Lagrange equations for the order
parameter profiles, $d^2 S/d y^2=2S-6S^2+4S^3$, are given by
\begin{equation}
S_{\pm}(y)=\frac{S_{\pm}(0)}{S_{\pm}(0)+(1-S_{\pm}(0))\exp(\mp\sqrt{2}y)}
\label{LdGprofiles}
\end{equation}
where $S_+(y)$ and $S_-(y)$ are, respectively,
the nematic order parameter profiles, as
a function of the distance to the planar substrate $y$, corresponding to a
bulk nematic or isotropic phase in bulk, and $S_\pm(0)$ are their values in
contact with the substrate.
The surface tensions between the planar
substrate and the nematic or isotropic phase are given by
\begin{eqnarray}
\sigma_{nw}&=&\int_0^\infty\left[\phi_{\mathrm{el}}(S_+)+\phi_{\mathrm{bulk}}(S_+)\right] dy-w S_+(0)
\nonumber\\&=&
\frac{\sqrt{2}}{6}(1+2S_+(0))(1-S_+(0))^2-wS_+(0)\label{snwLdG}\\
\sigma_{iw}&=&\int_0^\infty\left[\phi_{\mathrm{el}}(S_-)+\phi_{\mathrm{bulk}}(S_-)\right] dy-w S_-(0)
\nonumber\\&=&
\frac{\sqrt{2}}{6}(3-2S_-(0))S_-(0)^2-wS_-(0)\label{siwLdG}
\end{eqnarray}
The nematic order parameter values at contact with the substrate
can be obtained from the boundary condition $(dS_{\pm}/dy)(y=0)=-w$ as:
\begin{equation}
S_{\pm}(0)=\frac{1}
{1\mp\left(\frac{1}{\sqrt{2}w}\pm 1\right)+\sqrt{\left(\frac{1}{\sqrt{2}w}\pm 1
\right)^2-1}}
\label{defspm}
\end{equation}
The planar contact angle $\theta_\pi$ is obtained from Young's law
\begin{equation}
\cos\theta_\pi=\frac{\sigma_{iw}-\sigma_{nw}}{\sigma_{in}}
\label{costhetaLdG}
\end{equation}
where $\sigma_{IN}\approx 0.178$ for $\kappa=2$.

The roughness $r$ of a sinusoidal substrate, is given in
terms of the amplitude $A$ and the wavenumber $q$ as:
\begin{equation}
r=\frac{2E(-(qA)^2)}{\pi}
\label{roughnesssinusoidal}
\end{equation}
where $E(x)$ is the complete elliptic integral of second kind.

\begin{figure}[t]
\centerline{\includegraphics[width=.95\columnwidth]{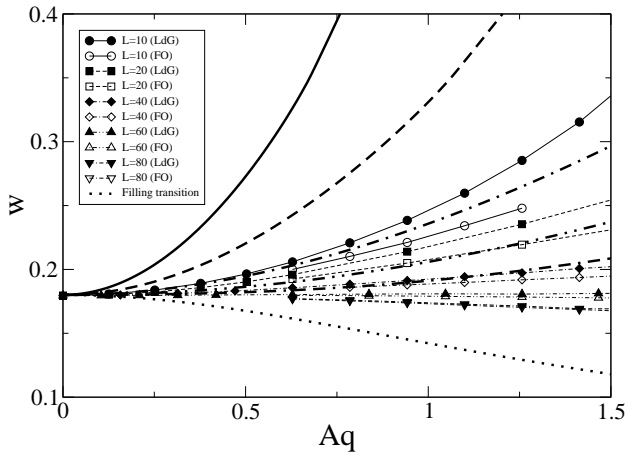}}
\caption {Location of the $D$-$W^\perp$ wetting transitions for
$\kappa=2$
and different values of $L/\xi=10$ (circles and solid lines), $20$ (squares and dashed lines), $40$ (rhombi and dot-dashed lines), $60$ (upward triangles and dashed-double dot lines) and $80$ (downward triangles and double dashed-dot lines). Symbols
correspond to the numerical results from full minimization of the LdG model
(filled symbols), and the FO model (open symbols).
Thick lines indicate the predictions from the modified Wenzel law,
Eq. (\ref{generalized_wenzel_law}), using Berreman's approach for the elastic
term. Thin lines only serve as guides for the
eyes for the numerical results using the same style as the thick lines for same parameter $L$. For comparison, the prediction from usual
Wenzel's law, Eq. (\ref{wenzel}), is shown (thick dotted line).
}
\label{fig6}
\end{figure}

Finally, for the elastic term we have to resort to some numerical
calculations. As a first estimate, we may use Berreman's approximation
$\Fcal_d\approx \Acal K(Aq)^2q/4$ \cite{Berreman_1972}.
Fig. \ref{fig6} shows the comparison between the
numerical and the modified Wenzel law predictions for
the $D$-$W^\perp$ transition line when Berreman's approach for the elastic
free-energy contribution is used. They are in qualitative agreement:
for small $L$ the wetting transition is shifted towards larger values of $w$,
and as $L$ increases, the transition line moves down, approaching
Wenzel's prediction, Eq. (\ref{wenzel}). A similar behaviour was observed
for the wetting transition of the saw-toothed substrate
\cite{Patricio_Pham_RomeroEnrique_2008,Patricio_Romero_etal_2011}.
However, there are quantitative discrepancies even for
the largest system considered. The explanation for these discrepancies is that
Berreman's expression is only accurate for small $Aq$ and $wL$ very large
(strong anchoring conditions), but it overestimates the exact result of
Frank theory \cite{Barbero_etal_2008}.
So, we have estimated the elastic energy obtained by using a Frank-Oseen (FO)
model with a Rapini-Papoular surface contribution, where the interaction
strength (which depends on $w$) is obtained from a contrained minimisation
of the LdG model in presence of a planar wall and a given nematic director
orientation at a distance $1.5\xi$ \cite{RomeroEnrique_Pham_Patricio_2010}.
We used a similar adaptative-mesh finite-element method to minimise
the FO model to that explained above for the LdG model. Our results
show that, for $Aq\sim 1$ and assuming strong anchoring
conditions, the elastic contribution is reduced, taking a value of around
$75\%$ of that obtained from Berreman's expression. However,
if weak anchoring is considered, a further reduction is observed, even for
$L=80\xi$. This fact contrasts with our previous results in the sawtoothed
substrate, in which the strong anchoring regime is reached for relatively
smaller values of $L$. When the numerically obtained values of
$\Fcal_d$ are used in Eq. (\ref{generalized_wenzel_law}), the predictions
for the wetting transition from the generalized Wenzel law are in very
good agreement with the full numerical values from the LdG model for $L\ge
60\xi$ (see Fig. \ref{fig6}).

\subsubsection{Case of $\kappa = 0$ or $\kappa < 0$}

\begin{figure}[t]
  \centerline{\includegraphics[width=0.95\columnwidth]{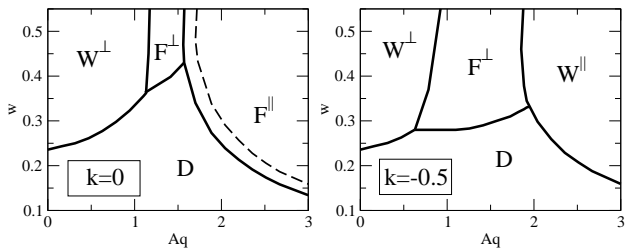}}
  \caption {Global phase diagram for $L = 10\xi$
and (left) $\kappa = 0$ ; (right) $\kappa = -1/2$. The dashed line on
the lefthandside figure is the $D-W^\parallel$ metastable transition
line.  }
\label{fig7}
\end{figure}

Now we turn to the effect of $\kappa$ in the adsorption phase
diagram. As it was already stated, for $\kappa=2$ the nematic fluid
has a preferred planar anchoring at the interface. As $\kappa$
decreases the free-energy cost to anchor the nematic in a parallel or
perpendicular orientation approach each other and become identical for
$\kappa=0$. If we further decrease $\kappa$, the system starts to
prefer a perpendicular orientation at the interface. As a consequence,
new interfacial states emerge, as the filled state $F^\perp$ (see
Fig. \ref{fig4}-(6)). These new states appear at intermediate values
of $Aq$ for large $w$, and the $NI$ interface shows a pronounced
curvature following the substrate shape. The global adsorption phase
diagrams for $L=10\xi$ are shown in Fig. \ref{fig7}. Again, the
continuous lines correspond to first-order transition lines which meet
at triple points. For $\kappa=0$, the region of the phase diagram
corresponding to $F^\perp$ states is located between the regions
corresponding to $W^\perp$ and $F^\parallel$ states, with almost
vertical boundaries between them. The $D$-$F^\perp$ transition is
weaker than the other transitions, so it is almost observed as a
crossover for $L=10\xi$. Finally, the $W^\parallel$ states free-energy
branch approach to the $F^\parallel$ free-energy branch. The dashed
line corresponds to the (metastable) $D$-$W^\parallel$ transition. For
$\kappa=-1/2$, the $W^\parallel$ states become the stable states for
large $Aq$ and $w$. So, the phase diagram is reminiscent of that for
$\kappa=0$, by swapping the $F^\parallel$ states with the
$W^\parallel$ states.  It is interesting to note that
it is possible to observe
reentrance of $F^\perp$ states as $w$ is increased: the fluid undergoes a
wetting transition from the $F^\perp$ state to the $W^\parallel$ state,
and for larger values of $w$ there is a reverse dewetting transition from
the $W^\parallel$  to the $F^\perp$ state. However, this
sequence of three transitions is limited to a very narrow range of
values of $Aq$.

\section{Conclusions}

In this article we have shown that the filling and wetting transition
sequence of an isotropic fluid on a sinusoidal substrate is deeply
changed for nematic liquid crystals.  Substrates favoring
perpendicular anchoring in contact with nematic liquid crystals, that
favor parallel anchoring at the nematic-isotropic interface, exhibit
only a wetting (and not filling) transition for small $Aq$, while for
large $Aq$ only filling (and not wetting) transitions occur.  In the
latter case, the complete wetting transition is prevented by the pinning of
the nematic-isotropic interface at the crests of the sinusoidal
substrate, thus hindering the nucleation of topological defects that
would otherwise be present in the wet configuration at large enough
anchoring strengths.  When perpendicular anchoring is favored at both
the substrate and the interface, the filling and wetting transition
lines change places.  In this case the wetting transition may occur
for larger $Aq$, whilst the filling transition occurs for intermediate
$Aq$. Finally, in conditions such that the anchoring at the
nematic-isotropic
interface may be either planar or homeotropic, the phase diagram shows mixed
features of the two previous cases. The analysis of these situations is far
more complex than in the previous cases, as any perturbation can change
the delicate balance between the different contributions to the free energy.

In order to check the validity of the phenomenological
generalized Wenzel's law, Eq. (\ref{generalized_wenzel_law}), we have
compared its predictions with the numerical transition values for the
one-step $D-W^\perp$ wetting transition. We see
that they are in excellent agreement
for $L\ge 60\xi$. This fact shows that
elastic effects are important in the range of values of wavelength $L$
considered in our work. In particular, they stabilize the $D$ state with
respect to the $W^\perp$ state,
as elastic deformations are a positive contribution
to the surface free energy only present in the latter state. On the other
hand, the location of the wetting transition approaches the value predicted
by the typical Wenzel's law, Eq. (\ref{wenzel}), when $L$ increases.

\acknowledgments

We acknowledge the support from FCT (Portugal) through Grant
No. PTDC/FIS/098254/2008, SFRH/BPD/40327/2007 (N. M. S.), Pluriannual
Contract with CFTC, and and Ac\c c\~ao Integrada Luso-Espanhola Ref. E
17/09 (P. P.).  J.M.R.-E. also acknowledges financial support from
Spanish Ministerio de Ciencia e Innovacion through Grants No.
FIS2009-09326 and HP2008-0028, and Junta de Andaluc\'{\i}a through
Grant No.  P09-FQM-4938.


\begin{thebibliography}{10}%
\makeatletter
\providecommand \@ifxundefined [1]{%
 \ifx #1\undefined \expandafter \@firstoftwo
 \else \expandafter \@secondoftwo
\fi
}%
\providecommand \@ifnum [1]{%
 \ifnum #1\expandafter \@firstoftwo
 \else \expandafter \@secondoftwo
\fi
}%
\providecommand \enquote [1]{``#1''}%
\providecommand \bibnamefont  [1]{#1}%
\providecommand \bibfnamefont [1]{#1}%
\providecommand \citenamefont [1]{#1}%
\providecommand\href[0]{\@sanitize\@href}%
\providecommand\@href[1]{\endgroup\@@startlink{#1}\endgroup\@@href}%
\providecommand\@@href[1]{#1\@@endlink}%
\providecommand \@sanitize [0]{\begingroup\catcode`\&12\catcode`\#12\relax}%
\@ifxundefined \pdfoutput {\@firstoftwo}{%
 \@ifnum{\z@=\pdfoutput}{\@firstoftwo}{\@secondoftwo}%
}{%
 \providecommand\@@startlink[1]{\leavevmode}%
 \providecommand\@@endlink[0]{}%
}{%
 \providecommand\@@startlink[1]{%
  \leavevmode
  \pdfstartlink
   attr{/Border[0 0 1 ]/H/I/C[0 1 1]}%
   user{/Subtype/Link/A<</Type/Action/S/URI/URI(#1)>>}%
  \relax
 }%
 \providecommand\@@endlink[0]{\pdfendlink}%
}%
\providecommand \url  [0]{\begingroup\@sanitize \@url }%
\providecommand \@url [1]{\endgroup\@href {#1}{\urlprefix}}%
\providecommand \urlprefix [0]{URL }%
\providecommand \Eprint[0]{\href }%
\@ifxundefined \urlstyle {%
  \providecommand \doi [1]{doi:\discretionary{}{}{}#1}%
}{%
  \providecommand \doi [0]{doi:\discretionary{}{}{}\begingroup
  \urlstyle{rm}\Url }%
}%
\providecommand \doibase [0]{http://dx.doi.org/}%
\providecommand \Doi[1]{\href{\doibase#1}}%
\providecommand \bibAnnote [3]{%
  \BibitemShut{#1}%
  \begin{quotation}\noindent
    \textsc{Key:}\ #2\\\textsc{Annotation:}\ #3%
  \end{quotation}%
}%
\providecommand \bibAnnoteFile [2]{%
  \IfFileExists{#2}{\bibAnnote {#1} {#2} {\input{#2}}}{}%
}%
\providecommand \typeout [0]{\immediate \write \m@ne }%
\providecommand \selectlanguage [0]{\@gobble}%
\providecommand \bibinfo [0]{\@secondoftwo}%
\providecommand \bibfield [0]{\@secondoftwo}%
\providecommand \translation [1]{[#1]}%
\providecommand \BibitemOpen[0]{}%
\providecommand \bibitemStop [0]{}%
\providecommand \bibitemNoStop [0]{.\EOS\space}%
\providecommand \EOS [0]{\spacefactor3000\relax}%
\providecommand \BibitemShut [1]{\csname bibitem#1\endcsname}%
\bibitem{Rascon_2000_2}%
  \BibitemOpen
  \bibfield{author}{%
  \bibinfo {author} {\bibfnamefont{C.}~\bibnamefont{Rasc\'on}}\ and\ \bibinfo
  {author} {\bibfnamefont{A.~O.}\ \bibnamefont{Parry}},\ }%
  \bibfield{journal}{%
  \bibinfo {journal} {Nature}\ }%
  \textbf{\bibinfo {volume} {407}},\ \bibinfo {pages} {986} (\bibinfo {year}
  {2000})%
  \bibAnnoteFile{NoStop}{Rascon_2000_2}%
\bibitem{Callies_Quere_2005}%
  \BibitemOpen
  \bibfield{author}{%
  \bibinfo {author} {\bibfnamefont{M.}~\bibnamefont{Calli\`es}}\ and\ \bibinfo
  {author} {\bibfnamefont{D.}~\bibnamefont{Qu\'er\'e}},\ }%
  \bibfield{journal}{%
  \bibinfo {journal} {Soft Matter}\ }%
  \textbf{\bibinfo {volume} {1}},\ \bibinfo {pages} {55} (\bibinfo {year}
  {2005})%
  \bibAnnoteFile{NoStop}{Callies_Quere_2005}%
\bibitem{Bonn_etal_2009}%
  \BibitemOpen
  \bibfield{author}{%
  \bibinfo {author} {\bibfnamefont{D.}~\bibnamefont{Bonn}}, \bibinfo {author}
  {\bibfnamefont{J.}~\bibnamefont{Eggers}}, \bibinfo {author}
  {\bibfnamefont{J.}~\bibnamefont{Indekeu}}, \bibinfo {author}
  {\bibfnamefont{J.}~\bibnamefont{Meunier}},\ and\ \bibinfo {author}
  {\bibfnamefont{E.}~\bibnamefont{Rolley}},\ }%
  \bibfield{journal}{%
  \bibinfo {journal} {Rev. Mod. Phys.}\ }%
  \textbf{\bibinfo {volume} {81}},\ \bibinfo {pages} {739} (\bibinfo {year}
  {2009})%
  \bibAnnoteFile{NoStop}{Bonn_etal_2009}%
\bibitem{Wenzel_1936}%
  \BibitemOpen
  \bibfield{author}{%
  \bibinfo {author} {\bibfnamefont{R.~N.}\ \bibnamefont{Wenzel}},\ }%
  \bibfield{journal}{%
  \bibinfo {journal} {Ind. Eng. Chem.}\ }%
  \textbf{\bibinfo {volume} {28}},\ \bibinfo {pages} {988} (\bibinfo {year}
  {1936})%
  \bibAnnoteFile{NoStop}{Wenzel_1936}%
\bibitem{Cassie_Baxter_1944}%
  \BibitemOpen
  \bibfield{author}{%
  \bibinfo {author} {\bibfnamefont{A.~B.~D.}\ \bibnamefont{Cassie}}\ and\
  \bibinfo {author} {\bibfnamefont{S.}~\bibnamefont{Baxter}},\ }%
  \bibfield{journal}{%
  \bibinfo {journal} {Trans. Faraday Soc.}\ }%
  \textbf{\bibinfo {volume} {40}},\ \bibinfo {pages} {546} (\bibinfo {year}
  {1944})%
  \bibAnnoteFile{NoStop}{Cassie_Baxter_1944}%
\bibitem{deGennes_1985}%
  \BibitemOpen
  \bibfield{author}{%
  \bibinfo {author} {\bibfnamefont{P.~G.}\ \bibnamefont{de~Gennes}},\ }%
  \bibfield{journal}{%
  \bibinfo {journal} {Rev. Mod. Phys.}\ }%
  \textbf{\bibinfo {volume} {57}},\ \bibinfo {pages} {827} (\bibinfo {year}
  {1985})%
  \bibAnnoteFile{NoStop}{deGennes_1985}%
\bibitem{Patricio_Pham_RomeroEnrique_2008}%
  \BibitemOpen
  \bibfield{author}{%
  \bibinfo {author} {\bibfnamefont{P.}~\bibnamefont{Patr\'icio}}, \bibinfo
  {author} {\bibfnamefont{C.-T.}\ \bibnamefont{Pham}},\ and\ \bibinfo {author}
  {\bibfnamefont{J.~M.}\ \bibnamefont{Romero-Enrique}},\ }%
  \bibfield{journal}{%
  \bibinfo {journal} {Eur. Phys. J. E.}\ }%
  \textbf{\bibinfo {volume} {26}},\ \bibinfo {pages} {97} (\bibinfo {year}
  {2008})%
  \bibAnnoteFile{NoStop}{Patricio_Pham_RomeroEnrique_2008}%
\bibitem{Swain_1998}%
  \BibitemOpen
  \bibfield{author}{%
  \bibinfo {author} {\bibfnamefont{P.~S.}\ \bibnamefont{Swain}}\ and\ \bibinfo
  {author} {\bibfnamefont{R.}~\bibnamefont{Lipowsky}},\ }%
  \bibfield{journal}{%
  \bibinfo {journal} {Langmuir}\ }%
  \textbf{\bibinfo {volume} {14}},\ \bibinfo {pages} {6772} (\bibinfo {year}
  {1998})%
  \bibAnnoteFile{NoStop}{Swain_1998}%
\bibitem{Rascon_1999}%
  \BibitemOpen
  \bibfield{author}{%
  \bibinfo {author} {\bibfnamefont{C.}~\bibnamefont{Rasc\'on}}, \bibinfo
  {author} {\bibfnamefont{A.~O.}\ \bibnamefont{Parry}},\ and\ \bibinfo {author}
  {\bibfnamefont{A.}~\bibnamefont{Sartori}},\ }%
  \bibfield{journal}{%
  \bibinfo {journal} {Phys. Rev. E}\ }%
  \textbf{\bibinfo {volume} {59}},\ \bibinfo {pages} {5697} (\bibinfo {year}
  {1999})%
  \bibAnnoteFile{NoStop}{Rascon_1999}%
\bibitem{Rejmer_2000}%
  \BibitemOpen
  \bibfield{author}{%
  \bibinfo {author} {\bibfnamefont{K.}~\bibnamefont{Rejmer}}\ and\ \bibinfo
  {author} {\bibfnamefont{M.}~\bibnamefont{Napi\'orkowski}},\ }%
  \bibfield{journal}{%
  \bibinfo {journal} {Phys. Rev. E}\ }%
  \textbf{\bibinfo {volume} {62}},\ \bibinfo {pages} {588} (\bibinfo {year}
  {2000})%
  \bibAnnoteFile{NoStop}{Rejmer_2000}%
\bibitem{Patricio_Romero_etal_2011}%
  \BibitemOpen
  \bibfield{author}{%
  \bibinfo {author} {\bibfnamefont{P.}~\bibnamefont{Patr\'icio}}, \bibinfo
  {author} {\bibfnamefont{J.~M.}\ \bibnamefont{Romero-Enrique}}, \bibinfo
  {author} {\bibfnamefont{N.~M.}\ \bibnamefont{Silvestre}}, \bibinfo {author}
  {\bibfnamefont{N.~R.}\ \bibnamefont{Bernardino}},\ and\ \bibinfo {author}
  {\bibfnamefont{M.~M.}\ \bibnamefont{{Telo~da~Gama}}},\ }%
  \bibfield{journal}{%
  \bibinfo {journal} {Mol. Phys.}\ }%
  \textbf{\bibinfo {volume} {109}},\ \bibinfo {pages} {1067} (\bibinfo {year}
  {2011})%
  \bibAnnoteFile{NoStop}{Patricio_Romero_etal_2011}%
\bibitem{Berreman_1972}%
  \BibitemOpen
  \bibfield{author}{%
  \bibinfo {author} {\bibfnamefont{D.~W.}\ \bibnamefont{Berreman}},\ }%
  \bibfield{journal}{%
  \bibinfo {journal} {Phys. Rev. Lett.}\ }%
  \textbf{\bibinfo {volume} {28}},\ \bibinfo {pages} {1683} (\bibinfo {year}
  {1972})%
  \bibAnnoteFile{NoStop}{Berreman_1972}%
\bibitem{RomeroEnrique_Pham_Patricio_2010}%
  \BibitemOpen
  \bibfield{author}{%
  \bibinfo {author} {\bibfnamefont{J.~M.}\ \bibnamefont{Romero-Enrique}},
  \bibinfo {author} {\bibfnamefont{C.-T.}\ \bibnamefont{Pham}},\ and\ \bibinfo
  {author} {\bibfnamefont{P.}~\bibnamefont{Patr\'icio}},\ }%
  \bibfield{journal}{%
  \bibinfo {journal} {Phys. Rev. E}\ }%
  \textbf{\bibinfo {volume} {82}},\ \bibinfo {pages} {011707} (\bibinfo {year}
  {2010})%
  \bibAnnoteFile{NoStop}{RomeroEnrique_Pham_Patricio_2010}%
\bibitem{deGennes_1970}%
  \BibitemOpen
  \bibfield{author}{%
  \bibinfo {author} {\bibfnamefont{P.-G.}\ \bibnamefont{de~Gennes}},\ }%
  \bibfield{journal}{%
  \bibinfo {journal} {Mol. Cryst. Liq. Cryst.}\ }%
  \textbf{\bibinfo {volume} {12}},\ \bibinfo {pages} {193} (\bibinfo {year}
  {1971})%
  \bibAnnoteFile{NoStop}{deGennes_1970}%
\bibitem{Patricio_Telo_Dietrich_2002}%
  \BibitemOpen
  \bibfield{author}{%
  \bibinfo {author} {\bibfnamefont{P.}~\bibnamefont{Patr\'icio}}, \bibinfo
  {author} {\bibfnamefont{M.~M.}\ \bibnamefont{{Telo~da~Gama}}},\ and\ \bibinfo
  {author} {\bibfnamefont{S.}~\bibnamefont{Dietrich}},\ }%
  \bibfield{journal}{%
  \bibinfo {journal} {Phys. Rev. Lett.}\ }%
  \textbf{\bibinfo {volume} {88}},\ \bibinfo {pages} {245502} (\bibinfo {year}
  {2002})%
  \bibAnnoteFile{NoStop}{Patricio_Telo_Dietrich_2002}%
\bibitem{deGennes_Prost_1995}%
  \BibitemOpen
  \bibfield{author}{%
  \bibinfo {author} {\bibfnamefont{P.~G.}\ \bibnamefont{de~Gennes}}\ and\
  \bibinfo {author} {\bibfnamefont{J.}~\bibnamefont{Prost}},\ }%
  \emph{\bibinfo {title} {The Physics of Liquid Crystals}},\ \bibinfo {edition}
  {2nd}\ ed.\ (\bibinfo {publisher} {Clarendon Press, Oxford},\ \bibinfo {year}
  {1995})%
  \bibAnnoteFile{NoStop}{deGennes_Prost_1995}%
\bibitem{Andrienko_Tasinkevytch_Patricio_etal_2004}%
  \BibitemOpen
  \bibfield{author}{%
  \bibinfo {author} {\bibfnamefont{D.}~\bibnamefont{Andrienko}}, \bibinfo
  {author} {\bibfnamefont{M.}~\bibnamefont{Tasinkevych}}, \bibinfo {author}
  {\bibfnamefont{P.}~\bibnamefont{Patr\'icio}},\ and\ \bibinfo {author}
  {\bibfnamefont{M.~M.}\ \bibnamefont{{Telo~da~Gama}}},\ }%
  \bibfield{journal}{%
  \bibinfo {journal} {Phys. Rev. E}\ }%
  \textbf{\bibinfo {volume} {69}},\ \bibinfo {pages} {021706} (\bibinfo {year}
  {2004})%
  \bibAnnoteFile{NoStop}{Andrienko_Tasinkevytch_Patricio_etal_2004}%
\bibitem{Sheng_1976}%
  \BibitemOpen
  \bibfield{author}{%
  \bibinfo {author} {\bibfnamefont{P.}~\bibnamefont{Sheng}},\ }%
  \bibfield{journal}{%
  \bibinfo {journal} {Phys. Rev. Lett.}\ }%
  \textbf{\bibinfo {volume} {37}},\ \bibinfo {pages} {1059} (\bibinfo {year}
  {1976})%
  \bibAnnoteFile{NoStop}{Sheng_1976}%
\bibitem{Sheng_1982}%
  \BibitemOpen
  \bibfield{author}{%
  \bibinfo {author} {\bibfnamefont{P.}~\bibnamefont{Sheng}},\ }%
  \bibfield{journal}{%
  \bibinfo {journal} {Phys. Rev. A}\ }%
  \textbf{\bibinfo {volume} {26}},\ \bibinfo {pages} {1610} (\bibinfo {year}
  {1982})%
  \bibAnnoteFile{NoStop}{Sheng_1982}%
\bibitem{Braun_1996}%
  \BibitemOpen
  \bibfield{author}{%
  \bibinfo {author} {\bibfnamefont{F.~N.}\ \bibnamefont{Braun}}, \bibinfo
  {author} {\bibfnamefont{T.~J.}\ \bibnamefont{Sluckin}},\ and\ \bibinfo
  {author} {\bibfnamefont{E.}~\bibnamefont{Velasco}},\ }%
  \bibfield{journal}{%
  \bibinfo {journal} {J. Phys.: Condens. Matter}\ }%
  \textbf{\bibinfo {volume} {8}},\ \bibinfo {pages} {2741} (\bibinfo {year}
  {1996})%
  \bibAnnoteFile{NoStop}{Braun_1996}%
\bibitem{Patricio_Tasinkevych_Telo_2002}%
  \BibitemOpen
  \bibfield{author}{%
  \bibinfo {author} {\bibfnamefont{P.}~\bibnamefont{Patr\'icio}}, \bibinfo
  {author} {\bibfnamefont{M.}~\bibnamefont{Tasinkevych}},\ and\ \bibinfo
  {author} {\bibfnamefont{M.~M.}\ \bibnamefont{{Telo~da~Gama}}},\ }%
  \bibfield{journal}{%
  \bibinfo {journal} {Eur. Phys. J. E}\ }%
  \textbf{\bibinfo {volume} {7}},\ \bibinfo {pages} {117} (\bibinfo {year}
  {2002})%
  \bibAnnoteFile{NoStop}{Patricio_Tasinkevych_Telo_2002}%
\bibitem{Barbero_etal_2008}%
  \BibitemOpen
  \bibfield{author}{%
  \bibinfo {author} {\bibfnamefont{G.}~\bibnamefont{Barbero}}, \bibinfo
  {author} {\bibfnamefont{A.~S.}\ \bibnamefont{Gliozzi}}, \bibinfo {author}
  {\bibfnamefont{M.}~\bibnamefont{Scalerandi}},\ and\ \bibinfo {author}
  {\bibfnamefont{L.~R.}\ \bibnamefont{Evangelista}},\ }%
  \bibfield{journal}{%
  \bibinfo {journal} {Phys. Rev. E}\ }%
  \textbf{\bibinfo {volume} {77}},\ \bibinfo {pages} {051703} (\bibinfo {year}
  {2008})%
  \bibAnnoteFile{NoStop}{Barbero_etal_2008}%
\end{thebibliography}
%

\end{document}